# Controlling Spontaneous Orientation Polarization in Organic Semiconductors - The Case of Phosphine Oxides

Albin Cakaj, Markus Schmid, Alexander Hofmann, and Wolfgang Brütting*

*Institute of Physics, University of Augsburg, 86135 Augsburg, Germany*

E-mail: wolfgang.bruetting@physik.uni-augsburg.de

## Abstract

Upon film growth by physical vapor deposition, the preferential orientation of polar organic molecules can result in a non-zero permanent dipole moment (PDM) alignment, causing a macroscopic film polarization. This effect, known as spontaneous orientation polarization (SOP), was studied in the case of different phosphine oxides. We investigate the control of SOP by molecular design and film-growth conditions. Our results show that using less polar phosphine oxides with just one phosphor-oxygen bond yields an exceptionally high degree of SOP with the so-called giant surface potential (slope) reaching more than $150\,\mathrm{mV\,nm^{-1}}$ in a neat BCPO film grown at room temperature. Additionally, by altering the evaporation rate and the substrate temperature, we are able to control the SOP magnitude over a broad range from 0 to almost $300\,\mathrm{mV\,nm^{-1}}$. Diluting BCPO in a nonpolar host enhances the PDM alignment only marginally, but combining temperature control together with dipolar doping can result in almost perfectly aligned molecules with more than 80 % of their PDMs standing upright on the substrate on average.

## Introduction

The dipolar nature of organic molecular semiconductors has played an important role in the development of organic light-emitting diodes (OLEDs) for display applications, even though the community has just recently started to develop a mechanistic understanding for the formation of spontaneous orientation polarization (SOP) – a phenomenon that arises from preferential alignment of the permanent electrical dipole moments (PDM) of such dipolar molecules.

Already in the very first report of efficient thin-film OLEDs by Tang and van Slyke, a dipolar molecule, viz. the green fluorescent dye $Alq_3$, was used as electron transport and light emitting material.[1] $Alq_3$ – or more specifically, its meridional isomer, which is prevalent in evaporated fims – has a strong electrical dipole moment of about 4.5 D. More than 10 years later, it was found that such prototypical bilayer OLEDs with a non-polar hole transport layer (HTL) and a polar electron transport layer (ETL) exhibit a non-uniform internal electric field distribution,[2,3] which could be explained by partial alignment of the PDMs of the polar material $Alq_3$[4,5] to form a macroscopic electric polarization. This happens spontaneously at the film surface while it is grown by physical vapor deposition (PVD), as sketched in Fig. 1. SOP is typically detected either by measuring a change of the device capacitance caused by polarization charges ($\sigma$) at the interface between the non-polar HTL and the polar ETL, e.g. by impedance spectroscopy (IS)[2] or the displacement current method (DCM),[3,6] or by Kelvin probe (KP),[4] which yields a surface potential ($V_S$) that is proportional to the film thickness and can reach macroscopic dimensions of several 10 volts for a few 100 nm thick films. Thus, SOP is also known as giant surface potential (GSP).[4,6]

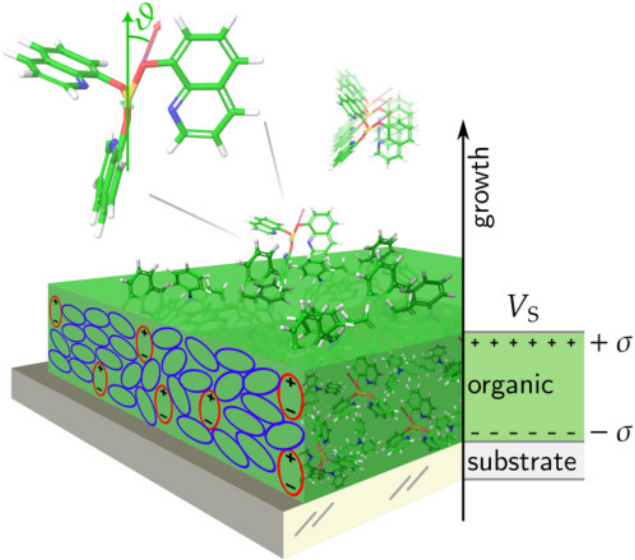

Figure 1: Schematic visualization of film growth by physical vapor deposition of organic molecules – here $Alq_3$ – with a permanent dipole moment (indicated as red arrow). If some of the molecules (shown as red ellipsoids) have preferential alignment in the direction perpendicular to the substrate plane, a macroscopic polarization is generated as evidenced by interfacial polarization charges ($\sigma$) of opposite sign at the bottom and top of the organic film and a surface potential ($V_S$) at its free surface to vacuum.

Meanwhile, quite a number of dipolar molecules used in OLEDs are known to exhibit SOP[7,8] and its implications for devices are intensively discussed.[9,10] For example, it was shown by drift-diffusion simulations[11] that the presence of negative interfacial charges between the HTL and $Alq_3$ in the prototypical bilayer OLED of Tang and van Slyke – or, equivalently, a positive GSP at the surface of the $Alq_3$ film – is favorable for electron injection. By contrast, if an ETL with negative GSP is used,[12] the injection current is significantly lower. On the other hand, later studies revealed that the presence of interfacial charge accumulation can cause exciton quenching[13] and that dipolar materials can have other negative effects on charge transport in a device,[14] which lead to the preliminary conclusion that SOP should be suppressed by either choosing non-polar materials or by growing dipolar materials at elevated substrate temperature – at about 80 to 90 % of their glass transition temperature $T_g$, where they form a so-called ultra-stable glass.[15,16] Another recent application of dipolar organic materials in vibrational power generators, however, would require to maximize their SOP in order to achieve the highest power output.[17]

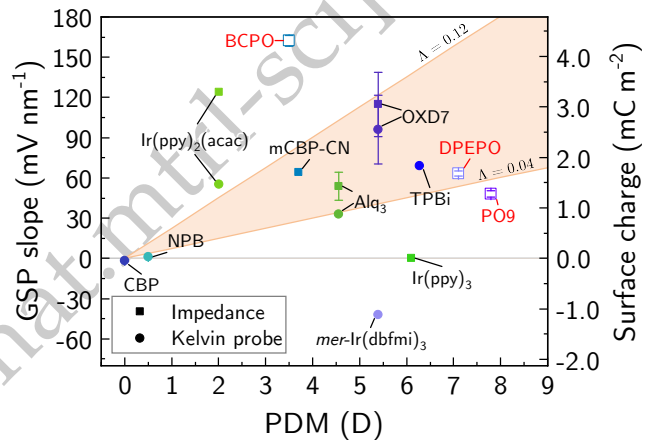

Figure 2: Overview of the polarization charge density $\sigma$ and GSP slope $m_{GSP}$ of various polar materials exhibiting SOP, which were studied in our lab over recent years, plotted vs. their permanent dipole moment (PDM). A detailed list of the materials and the respective references is given in the SI, Section S1. Three phosphine oxides investigated with IS in a previous publication are given in red color.[18] The data was aquired using either impedance spectroscopy (squares) or Kelvin probe (circles). The relation between the two ordinate axes is fixed with $\varepsilon_r = 3$.

From this short overview, it is apparent that SOP has multi-faceted implications for organic electronic devices, requiring a deeper mechanistic understanding of how to control it and, thus, tune it toward the required direction. In general, SOP (and the concomitant GSP) are proportional to the PDM of the dipolar molecular species under considera-tion. This is shown in Fig. 2 for a compilation of materials studied in our lab over recent years (using either IS or KP) – including three of the four materials studied here. For similar graphs with additional materials we refer to the literature.[7,8,10,19]

A quantitative analysis (details are given in Section S1 of the Supporting Information) shows that the degree of alignment of molecular PDMs in the direction perpendicular to the substrate plane is quite low. Considering that the theoretical upper limit of the GSP slope (per Debye) is about $190\,\mathrm{mV\,nm^{-1}\,D^{-1}}$ – see the SI, one can estimate the corresponding degree of dipole alignment to be between about 4 % and 12 % only for the shaded area in Fig. 2, similar to other publications.[8,20–22] This is not unexpected because of the amorphous ("glassy") nature of these films with almost random orientation of the molecules. Another factor is electrical dipole-dipole interactions, which would tend to align the PDMs of neighboring molecules in an anti-parallel fashion, thus canceling the overall net dipole moment of the pair. Nevertheless, there seem to be some "outliers" in Fig. 2 that do not fall into the shaded area. One is BCPO, which will be discussed in further detail in this work. Another prominent example is Ir(ppy)$_2$(acac), which has been found to simultaneously exhibit high degree of alignment for both its PDM (standing upright) and transition dipole moment (TDM, lying down).[20] This is caused by its special molecular geometry with PDM and TDM being mutually orthogonal to each other.

Recent studies on molecular orientation in the context of their optical properties – most prominently, their transition dipole moments which are a very powerful handle to improve light outcoupling for OLEDs[23] – have shown different ways to control their alignment. These "design rules" include anisotropic molecular shapes,[24] control of the substrate temperature and evaporation rate during deposition[16] or choice of high-$T_g$ host materials in light-emitting guest-host systems.[18,25] In the context of SOP, dilution of the dipolar species in a non-polar host, i.e. dipo-

lar doping, has also been shown to affect the degree of alignment.[20,21]

Quantitative prediction of SOP in evaporated thin films is still challenging because of the complexity of the systems having, for example, many different conformers or isomers and the presence of strong polarization effects in solid films, which make numerical calculations very cost intensive.[26,27] But one may expect that cost-efficient methods to predict the optical TDM orientation[28] may in some way be transferable for SOP alignment, too. Finally, we want to mention that a dedicated design of molecules with the aim to control the magnitude and direction of SOP has just recently started.[19,29]

Here we investigate a series of molecules with phosphine oxide (PO) groups, which are frequently used as host materials and electron transporting/hole blocking layers in phosphorescent and thermally activated delayed fluorescent (TADF) OLEDs, because they have large energy gaps paired with high triplet energies.[30–32] And, in the context of TADF emitters, it is important that they have a non-vanishing electrical PDM which is required for enabling the reverse intersystem crossing process.[33,34] However, their SOP has not been studied systematically yet. Importantly, one material from this class (BCPO, as mentioned above) has been found to exhibit a record-high GSP of more than $150\,\mathrm{mV\,nm^{-1}}$, which is by a factor of 2 to 4 higher than other prototypical SOP materials, like Alq$_3$ or TPBi, having similar or even higher PDM, as shown in Fig. 2.

First, we compare four of these phosphine oxides, where two of them have just one PO group while the other two have two of them, which roughly doubles their PDM. Surprisingly, however, the obtained SOP (respectively, the GSP) is much larger in the former case of molecules with one PO group only. This indicates that the reason for high GSP is neither the mere presence of a strongly polar bond, like the P=O bond, nor the overall magnitude of the PDM of a molecule. Through a systematic variation of the film growth conditions for the champion material BCPO, we are able to control its SOP over a wide range yielding unprecedented GSP values of almost $300\,mV\,nm^{-1}$ for films grown on cooled substrates. Furthermore, we correlate electrical measurements of the PDM alignment with optical measurements of its TDM alignment and its birefringence in films to obtain additional information on molecular orientation. This is complemented by quantum chemical calculations to yield a comprehensive picture of the driving forces controlling the formation of SOP in these materials.

## Materials

For our experimental studies, we investigated four different phosphine oxides each characterized by at least one strongly polar P=O double bond and three organic groups attached to the phosphor atom. The molecular structures can be seen Figure 3. The investigated molecules have in general the structure $R^1R^2R^3P{=}O$. The strong electronegative oxygen leads to a large PDM in many molecules of this material class. One could even speculate, that the P=O bond is so polar that it has ionic character, which causes additional molecular interactions in the film formation process.

Molecules of this material group are particularly used in OLEDs as host for deep blue emitters[35–39] and as electron tranport material due to their high triplet energy levels.[30–32] We studied four molecules from this class of materials: Bis[2-(diphenylphosphino)phenyl] ether oxide (DPEPO), 3,6-bis(diphenylphosphory(-)9-phenylcarbazole (PO9), phenyldi(pyren-2-yl)phosphine oxide (POPy2) and bis-4-(N-carbazo(yl)phenyl)pheny phosphine oxide (BCPO); their chemical structures are shown in Fig. 3. While BCPO and POPy2 are characterized by one P=O double bond, DPEPO and PO9 contain two of them. Side groups attached to the phosphor by single bonds can rotate with respect to the core structure and result in different energetically favorable configurations. Additionally, 4,4´-Bis(N-carbazolyl)-1,1´byphenyl (CBP) was used as a non-polar host for diluting BCPO. CBP is widely used as host material for fluorescent and phosphorescent emitters in OLEDs and as hole transport material.[40–44]

By using a combination of electrical and optical measurements we are able to probe the film average of the permanent dipole moment, the molecular polarizability and the transition dipole moment orientation. For this reason we calculated the molecular quantities for the optimized geometries. Due to the high molecular flexibility, caused by the rotational freedom of the single bonds attached to the phosphor atom, all phosphine oxides have multiple stable conformers, some of which will be discussed in the course of this publication. In order to obtain the microscopic properties of each specific molecule, density functional theory (DFT), time-dependent DFT and molecular dynamics (MD) calculations were performed. For details we refer to the methods section and the SI.

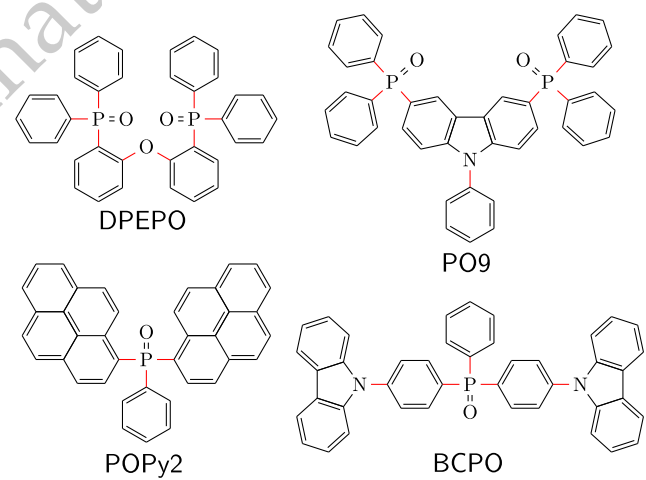

Figure 3: Chemical structures of the investigated phosphine oxides.

## Results

### Comparison of different phosphine oxides

All of the investigated phosphine oxide molecules are polar and harbor a permanent dipole moment originating from the P=O

bond. Thus, upon PVD their PDMs can exhibit an out-of-plane net alignment which leads to a macroscopic film polarization. This effect is called spontaneous orientation polarization and gives rise to charge accumulation in the adjacent layer of a device even below the device turn-on voltage. The responsible polarization charges ($\sigma$) can be determined by impedance spectroscopy (IS) and are a measure for the strength of SOP:

$$\sigma = p \left\langle \cos \vartheta_{PDM} \right\rangle n, \qquad (1)$$

wherein $p$ is the magnitude of the molecular PDMs, $\langle \cos \vartheta_{PDM} \rangle$ their average alignment along the surface normal of the film and $n$ their number density in the film. Note that $\sigma$ is positive on the surface of the film, if the net PDM alignment is pointing away from the substrate as shown in Fig. 1, which is typical for many SOP materials having positive GSP slope.

Accordingly, impedance spectroscopy on simple OLED stacks containing these phosphine oxides were performed. Details regarding the method and exemplary raw measurement data are given in the SI, Section S2. The obtained surface charge densities of the four phosphine oxides are plotted in Figure 4. They form two groups: DPEPO and PO9 with two PO groups each have similar $\sigma$ of 1.68 mC m$^{-2}$ and 1.28 mC m$^{-2}$, which is comparable to other polar molecules with PDMs in the range 7 D to 8 D.[8,45] But, surprisingly, POPy2 and BCPO exhibit significantly higher polarization charge densities of 3.43 mC m$^{-2}$ and 4.31 mC m$^{-2}$, respectively, even though they only have one polar PO bond each.

For better comparison we calculated the GSP slopes $m_{GSP}$ from the surface charge densities according to

$$m_{GSP} = \frac{\sigma}{\varepsilon_0 \varepsilon_r} = p \left\langle \cos \vartheta_{PDM} \right\rangle \frac{n}{\varepsilon_0 \varepsilon_r}, \qquad (2)$$

where the relative permittivity $\varepsilon_r$ of the specific material has to be known. This was experimentally determined for BCPO and POPy2 by IS measurements under variation of the ETL layer thickness, see the SI section S2 for details. The obtained values are (3.21 ± 0.15) for BCPO and (3.13 ± 0.14) for POPy2. For simplicity, however, a value of $\varepsilon_r = 3.0$ was used for all molecules in Figure 4. The resulting GSP slopes are compiled in Table 1. Details regarding the PDM calculation of the investigated phosphine oxides can be found in the discussion and the methods section. DPEPO and PO9 show a $m_{GSP}$ of 63.3 mV nm$^{-1}$ and 48.3 mV nm$^{-1}$, respectively, whereas BCPO and POPy2 exhibit significantly larger values of 151 mV nm$^{-1}$ and 123 mV nm$^{-1}$.

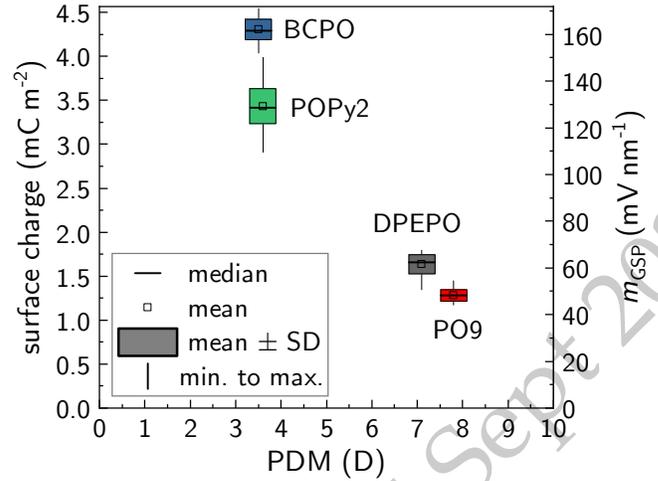

Figure 4: Measured surface charge and the calculated GSP slope values plotted against their respective PDMs for the investigated phosphine oxides. For simplicity, a dielectric constant of 3.0 was taken for all materials.

As already said in the introduction, the GSP slope is related to the PDM magnitude and, consequently, one would simply expect organic molecules with a larger PDM to harbor a bigger $m_{GSP}$, see Eq. 2. Many reported polar molecules show such a tendency including DPEPO and PO9 (see Fig. 2). However, it is clearly visible that BCPO and POPy2 deviate significantly. Thus, to cause such a high $m_{GSP}$, the degree of out-of-plane PDM alignment has to be much larger than in other molecules with a similar sized PDM.

The degree of alignment is given by the order parameter $\Lambda$:

$$\Lambda := \left\langle \cos \vartheta_{PDM} \right\rangle = \frac{m_{GSP}}{(pn/\varepsilon_0 \varepsilon_r)} = \frac{\sigma}{pn} \qquad (3)$$

Table 1: Investigated phosphine oxides and their properties.

| Material | PDM (D) | $\sigma$ (mC m$^{-2}$) | $m_{GSP}$ (mV nm$^{-1}$) | $\Lambda$ (%) | $T_g$ (°C) |
|---|---|---|---|---|---|
| PO9 | 7.8 | $1.28 \pm 0.07$ | 48.3 | 4.5 | 122 |
| DPEPO | 7.1 | $1.68 \pm 0.05$ | 63.3 | 5.6 | 93 |
| POPy2 | 3.6 | $3.43 \pm 0.20$ | 123 | 21 | NA |
| BCPO | 3.5 | $4.31 \pm 0.11$ | 151 | 32 | 137 |

While the degree of alignment for DPEPO and PO9 of 5.6 % and 4.5 % is comparable to other known organic molecules,[18] BCPO and POPy2 exhibited an extraordinary large degree of PDM alignment of 32 % and 21 %, respectively. Comparable degrees of PDM alignment were only achieved before in a neat film of the phosphorescent emitter Ir(ppy)$_2$(acac) or by dilution of the polar species in an non-polar host.[20,21,45]

Also included in Table 1 are the glass transition temperatures of the four materials, as far as they are available in the literature. As will be discussed later, they do not seem to be correlated to the observed magnitudes of the GSP slope or the degrees of alignment.

SOP formation is a consequence of the preferred alignment of molecular PDMs. However, since the PDM vectors usually have a specific direction on the molecular frame, it would potentially imply that the molecules themselves are aligned. This can be investigated by other techniques, like variable angle spectroscopic ellipsometry (VASE). With this technique we can obtain the wavelength dependent optical constants ($n(\lambda)$ & $k(\lambda)$) of neat organic films. From these we gain information regarding the average orientation of two other microscopic molecular properties with respect to the surface normal, namely the transition dipole moment (TDM) in absorption and the fundamental polarizability eigenvectors.

Organic films with uniaxial anisotropy can be described by two sets of optical constants, in-plane ("ordinary") and out-of-plane ("extraordinary"). A noticeable difference in the refractive indices ($n_o$, $n_e$) or the extinction coefficients ($k_o$, $k_e$) indicates optical anisotropy, i.e. birefringence, of the film and therefore a preferred molecular alignment. The raw data and the resulting wavelength dependent optical constants for the investigated phosphine oxides are shown in the supporting information, Section S3. All four materials show almost no difference between the in-plane and out-of-plane direction. Consequently the investigated films can be considered as optically isotropic – in spite of their SOP.

## Controlling the SOP of BCPO

It was previously reported that the film morphology of vapor-deposited organic molecules can be controlled by the evaporation conditions of the film, especially, the used evaporation rates and the substrate temperature $T_S$ in relation to the glass transition temperature $T_g$ of a material.[46] While previous studies focused on optical measurement techniques to investigate molecular orientation,[47–49] only few looked at SOP forming materials and their respective GSP slope $m_{GSP}$ under different film-growth conditions.[13,50]

Thus, to gain insight into the SOP formation process the influence of different evaporation conditions on the GSP slope were examined. For that purpose BCPO was chosen as the model candidate due to its exceptionally high $m_{GSP}$ in evaporated thin films. We grew films at varied substrate temperature $T_S$ over a range of 180 °C using a constant deposition rate of 0.3 Å s$^{-1}$ and measured their GSP in-situ with a Kelvin probe setup without breaking the vacuum; see the SI, Section S4.1 for the raw data. $m_{GSP}$ was obtained as the slope of the linear increasing contact potential difference (CPD) with increasing layer thicknesses. For the whole process of deposition and measuring, each sample was kept at its respective substrate temperature $T_S$. In Figure 5a the resulting $m_{GSP}$ values are plotted vs. the temperature ratio $T_S/T_g$ of BCPO, which has a $T_g = 137\,°C$.[35] By looking at the temperature dependence a clear trend can be seen. An increase in substrate temperature beyond room temperature leads to a decrease in SOP, while close to the glass transition temperature (at $T_S \approx 0.9T_g$) the $m_{GSP}$ vanishes. On the other side, cooling the substrate increases the degree of SOP significantly and a maximum of almost $300\,mV\,nm^{-1}$ can be achieved at $T_S = -70\,°C$. Compared to the usually used substrate temperatures around room temperature, the $m_{GSP}$ has obviously doubled its value.

Please note that in this measurement at room temperature, the BCPO film showed an $m_{GSP}$ of $145\,mV\,nm^{-1}$, which is lower compared to the value derived from the IS measurement and a previously reported value.[18] However, the samples for KP and IS measurements were prepared in different deposition chambers, and other organic molecules also showed differences in their respective $m_{GSP}$ values due to differences in the evaporation conditions.[10] We think that specifically different evaporation rates may play an important role, as will be discussed below.

Furthermore, the resulting PDM alignment parameter $\Lambda$ (Eq. 3) is plotted in Figure 5a as well. At the maximum GSP value, $62\,\%$ of all permanent dipole moments of BCPO are on average aligned along the surface normal. Compared to other common polar organic molecules prepared as neat films at room temperature, the PDM orientation of BCPO is apparently about 10 times larger.[18] Only by diluting polar molecules in a non-polar host could a similarly high degree of alignment be achieved, however, at a much lower total GSP.[20,21,45]

For further insights regarding a possible change in the optical anisotropy with changing the substrate temperature, post-growth VASE measurements were performed on the same samples at ambient conditions. The resulting substrate temperature dependent birefringence $\Delta n$ can be seen in Figure 5b; raw data obtained by VASE are shown in the SI, Section S5.1. Birefringence is the difference between the out-of-plane and in-plane component of the refractive index $\Delta n = n_e - n_o$ and contains information regarding the electronic polarizability in the film with respect to the surface normal. At room temperature, BCPO shows a slightly negative value of $\Delta n = -0.017$ corresponding to an optically almost isotropic film. Still the temperature dependent $\Delta n$ values show a trend comparable to previously investigated rod and disc-shaped organic molecules.[47–49] By heating the substrate above room temperature, the already small negative $\Delta n$ value approaches zero, while being even slightly positive at $T_S = 60\,°C$; this is equivalent to a very weak out-of-plane orientation. On the other hand, cooling the substrate causes the birefringence to decrease significantly. The increasingly larger negative values correspond to a tendency of in-plane alignment of the polarizability ellipsoid, but compared to previously studied molecules in a similar temperature range $(T_S/T_g)$[47] the overall magnitude of $\Delta n$ at all temperatures is small.

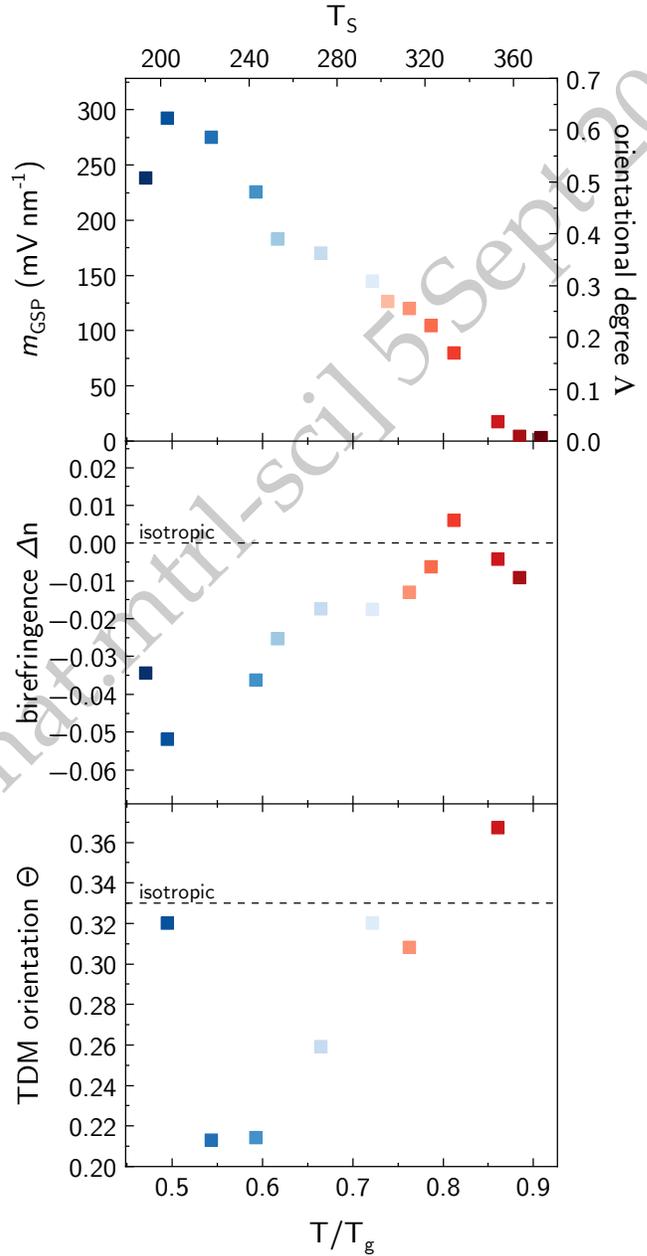

Figure 5: a) GSP slopes of BCPO films grown at different substrate temperature $T_S$ (upper axis) and the respective $T_S/T_g$ (lower axis). b) Birefringence $\Delta n$ of these BCPO films obtained from VASE and c) their TDM orientation. The raw data can be found in the SI, Sections S4.1, S5.1 & S6.

Molecular alignment can be further probed by performing angular dependent photoluminescence (ADPL) measurements. With this technique the average transition dipole moment orientation $\Theta = \langle \cos^2 \vartheta_{TDM} \rangle$ can be obtained. A value of 0.33 implies an isotropic TDM orientation, while a smaller or larger value corresponds to a preferential horizontal/vertical TDM alignment. The resulting values for BCPO prepared at different substrate temperatures can be seen in Figure 5c; raw data are in Section S6 of the SI. While at room temperature the alignment is close to isotropic, cooling increases the in-plane alignment of the transition dipole moments; vice versa heating causes a small out-of-plane orientation. This behavior is similar to the temperature dependent birefringence having a tendency for higher in-plane alignment at low temperatures, while showing almost no orientation at elevated temperatures. Remarkable is, that in all three measurements a loss in orientation at the lowest possible substrate temperature $T_S$ can be seen, which could indicate that at very low temperature surface diffusivity of evaporated molecules is too low to find suitable sites for favorable alignment.

Another control parameter in the process of deposition is the evaporation rate. Both the substrate temperature and the evaporation rate have a similar effect on the molecular packing.[51] Accordingly, BCPO films at different evaporation rates ranging from $0.05\,\text{Å}\,\text{s}^{-1}$ to $2.4\,\text{Å}\,\text{s}^{-1}$ were grown on substrates at room temperature and investigated via Kelvin probe (see SI, Section S4.2). Following from these measurements the $m_{GSP}$ values are plotted vs. their respective deposition rate in Figure 6. Increasing the evaporation rate causes an increase in the measured GSP slope $m_{GSP}$ and, vice versa, a smaller rate leads to a decrease. But over the whole chosen range of evaporation rates only a moderate change in the GSP slope by less than $20\,\text{mV}\,\text{nm}^{-1}$ can be observed.

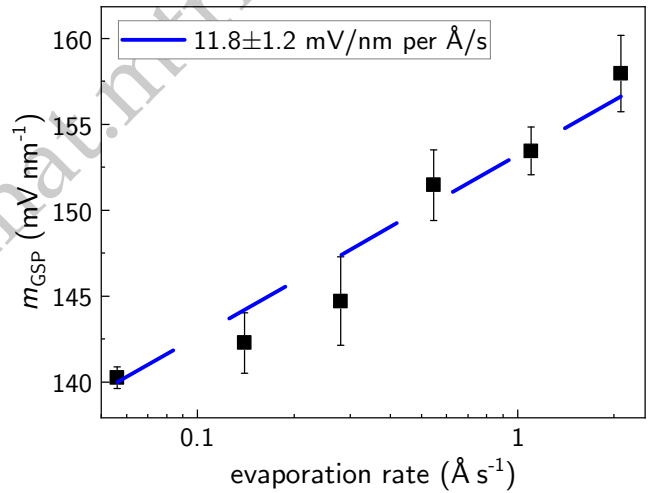

Figure 6: GSP slopes of BCPO films prepared with different evaporation rates at room temperature. The raw KP data are shown in the SI, Section S4.2.

It is known that changing the evaporation rate by a factor of 50 has almost the same impact on $m_{GSP}$ as a change of $10\,\text{K}$ in substrate temperature.[52] Thus, we can tentatively assign the differences in interfacial charge density and GSP slope between IS and KP measurements, which we mentioned above, to the use of different evaporation rates and/or slightly different $T_S$ in the respective deposition chambers.

Again VASE measurements were performed

for the samples prepared at different evaporation rates. The resulting birefringence values can be found in the SI, Section S5.2. All samples showed a slightly negative value comparable to the previously determined value at room temperature and an evaporation rate of $0.3\,\text{Å}\,\text{s}^{-1}$. Hence changing the evaporation rate affected the birefringence only marginally.

An additional approach to control SOP is dilution of the polar species in a non-polar host. Some material combinations exhibit a large increase in their PDM alignment depending on the doping concentration.[21,53] It was suggested that, due to the increased mutual distance and consequently reduced dipole-dipole interactions, the PDMs tend to form less antiparallel pairs. Therefore, BCPO was coevaporated at different volume concentration in non-polar CBP to form a guest-host system. The corresponding KP data are contained in the SI, Section S4.3. Note that CBP has a PDM of $0\,\text{D}$ and consequently does not exhibit SOP, as also shown in the SI.

Due to changes in molecular interactions, dilution can cause an increase in PDM alignment. The degree of enhancement in PDM alignment $\gamma(x)$ can be calculated as a function of doping concentration $x$:

$$\gamma(x) = \frac{\langle \cos \vartheta_{\text{mix}} \rangle}{\langle \cos \vartheta_{\text{neat}} \rangle} = \frac{m_{\text{GSP}_{\text{mix}}}(x) \cdot \varepsilon_{\text{mix}}}{m_{\text{GSP}_{\text{neat}}} \cdot \varepsilon_{\text{neat}} \cdot x} \quad (4)$$

with $\langle \cos \vartheta_{\text{mix}} \rangle$ and $\langle \cos \vartheta_{\text{neat}} \rangle$ being the average alignment values between the PDM direction and the surface normal in the mixed guest-host system and neat guest-only film. $m_{\text{GSP}_{\text{mix}}}(x)/m_{\text{GSP}_{\text{neat}}}$ and $\varepsilon_{\text{mix}}(x)/\varepsilon_{\text{neat}}$ are the GSP values and the dielectric constants of the mixture and the neat guest film, respectively. The dielectric constant of the mixed film can be estimated by an effective-medium approximation as

$$\varepsilon_{\text{mix}}(x) = \varepsilon_{\text{neat}} \cdot x + \varepsilon_{\text{host}} \cdot (1 - x) \ . \quad (5)$$

The resulting enhancement factor $\gamma(x)$ is plotted vs. the respective doping concentration $x$ in Figure 7. For all doping concentrations an increase in PDM alignment can be seen. But, overall, the enhancement is moderate and the maximum value corresponds to an increase of around $35\,\%$ compared to the non-diluted neat BCPO film.

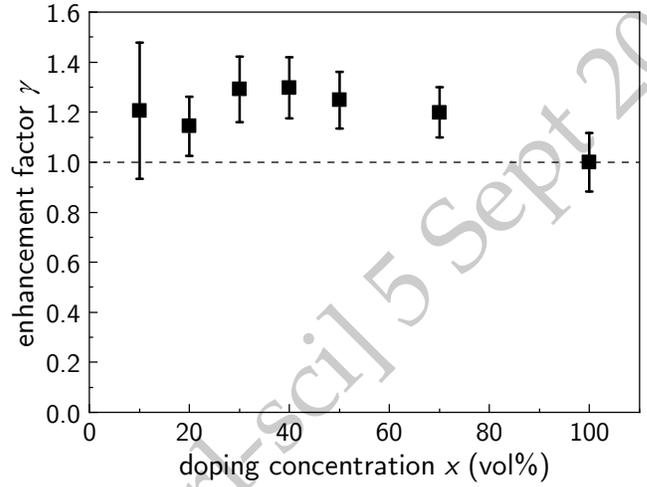

Figure 7: Enhancement factors of BCPO diluted in a non-polar CBP host. The raw KP data are shown in the SI, Section S4.3.

Finally, a combination of possible control methods should lead to even better PDM alignment. Due to the evaporation rate affecting the $m_{\text{GSP}}$ only moderately, temperature control and doping were combined. Thus, films consisting of $30\,\%$ BCPO in CBP were grown at low substrate temperatures. In that way, an alignment factor $\Lambda$ of $84\,\%$ (weighted by the volume percentage) was reached at $T_{\text{S}} = -50\,°\text{C}$; see SI, Section S4.3 for the raw KP data. Consequently almost all BCPO molecules are aligned and have their PDMs standing upright on the substrate in this guest-host system. Similar to the neat BCPO films, further cooling to $T_{\text{S}} = -70\,°\text{C}$ showed a decrease in $m_{\text{GSP}}$ and consequently PDM alignment, however, not as strong as for the neat film case.

## Discussion

To understand the influence of molecular structure on SOP formation, density functional theory (DFT) calculations were performed. Details regarding the procedure can be found in the methods section. The optimization of molecular structures resulted in many different stable conformers for each investigated molecule. Exemplary ones are shown in Figure 8a and in the SI, Section S7. The large number of different possible geometries is a consequence of the high molecular flexibility. Each phosphine oxide unit has three freely rotatable groups attached by single bonds to the phosphor atom, written formally as $R^1R^2R^3P{=}O$. Thereupon the inherent polarizability tensor can be significantly different for each stable conformer. In the end, averaging over all possible geometries in the film, each characterized by a different polarizability ellipsoid, can result in an optically isotropic film, even if the PO groups of the molecules in the film are highly aligned.

By contrast, the permanent dipole moment is primarily given by the relative direction of the P=O bond(s) and, thus, more or less independent of the other three single bonds at each phosphor atom. For BCPO and POPy2 with only one PO group the PDM magnitude is nearly constant for each calculated configuration with around 3.5 D (see SI, Section S7). For DPEPO and PO9, however, the PDM direction depends on the relative orientation of the two PO groups (Figure 8a). This results in a broad variety of conformers with different overall PDM. Moreover, they have twice the amount of freely rotatable bonds. For that reason molecular dynamics simulations were performed for DPEPO and PO9 to evaluate their respective PDM distributions. The result is shown in Figure 8b. By averaging over all populated conformers the average PDMs of DPEPO and PO9 are obtained as 7.1 D and 7.8 D, respectively, as already given in Table 1. Previously published values from DFT range from 2.2 D to 8.1 D for DPEPO[39,54] and 6.7 D for PO9.[18]

From these simulations we can conclude that a large number of possible molecular conformations can have a negative influence on molecular alignment in the film. This was previously reported by Yokoyama et al.[55] Additionally dipole-dipole interactions are known to be limiting the degree of SOP, because of antiparallel aligned PDMs being energetically more favorable.[55] For molecules with large PDM, like PO9 and DPEPO, this effect is more pronounced and leads to less preferential alignment and a smaller overall alignment parameter $\Lambda$ parameter compared to BCPO and POPy2.

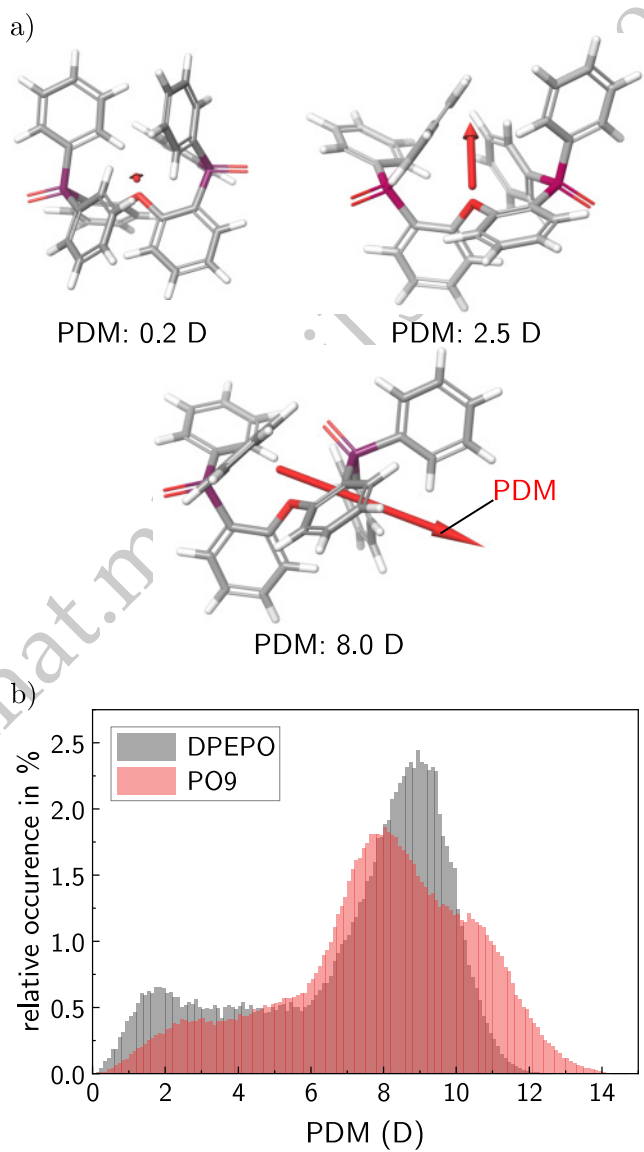

Figure 8: a) Exemplary conformers of DPEPO with different permanent dipole moments. b) Calculated PDM distributions of DPEPO and PO9 from molecular dynamics. The weighted average yields PDM values of 7.1 D for DPEPO and 7.8 D for PO9, respectively.

An additional argument comes from the electrostatic surface potential (ESP) of the four phosphine oxides, as shown in Figure 9a, where we find a large negative value at the position of the oxygen in the PO group. At the same time the PDM vector of BCPO and POPy2 lies along the P=O bond and points from the oxygen to the phosphor atom. From the results of IS and KP we can thus conclude that the oxygen atoms are aligning favorably towards the substrate. Previous studies on Ir-complexes concluded that a large anisotropy of the ESP is advantageous for molecular alignment,[56,57] because the close proximity of high and low ESP tend to form attractive forces for evaporated molecules arriving at the film surface (as indicated in Fig. 9b).

In the case of phosphine oxides hydrogen bonds and inductive van-der-Waals forces (Keesom forces) are possible. Hydrogen bonds can be formed between the free electron pairs of the oxygen atom in the PO group and the C-H groups from the underlying organic film. A similar mechanism plays a role in the alignment of transport materials and TADF emitters.[24,55,58,59] The attractive hydrogen bonds could be even experimentally proven by Yokoyama et al.[55] Additionally, the PO group is highly polar, which can induce a temporary dipole in a polarizable molecule nearby. These dipoles can then interact via attractive van-der-Waals forces. Both of these interactions can be considered as driving forces for aligning the PO group vertically on the substrate with the concomitant PDM pointing upward (Fig. 9b).

In the case of BCPO and POPy2, there is only one point of interaction to the underlying organic groups at the film surface, which leads to strong alignment. By contrast, since DPEPO an PO9 have two phosphine oxide groups, two rivaling interaction points are present.

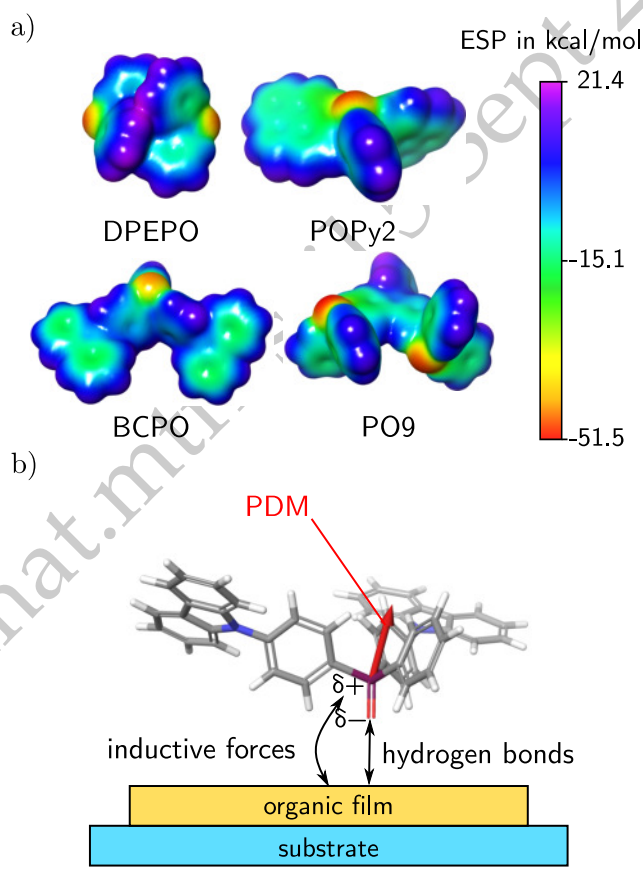

Figure 9: a) Calculated electrostatic surface potential (ESP) maps of the investigated phosphine oxides. A strong negative pole at the positions of the oxygen atoms can be seen. b) Schematic orientation mechanism of BCPO causing a high degree of SOP by forming hydrogen bonds and van-der-Waals interactions between a molecule arriving at the film surface and the underlying organic film.

Due to the broad distribution of conformers no preferred direction for alignment exists, thus, leading to an overall small $\Lambda$ value compared to BCPO and POPy2. This means that less PO groups, i.e. a smaller molecular PDM, is actually more favorable for achieving strong PDM alignment and, with that, a large SOP (or GSP slope).

We note that Tanaka et al. have recently published a unique design rule for controlling alignment and the direction of SOP independently.[19] They were able to achieve very strong alignment of molecules by minimizing the surface free energy through perfluorinated groups. At the same time, they could control the sign of the GSP (i.e. the direction of SOP) by using additional substituents with PDMs pointing either up- or downward on the substrate surface. In this way, they could achieve both positive and negative GSP slopes of similar magnitude as BCPO and POPy2.

Finally, we would also like to comment on the role of the glass transition temperatures $T_g$ for orientation of the studied four phosphine oxides (see Table 1). According to the surface equilibration mechanism put forward by Ediger et al., the growth of films exhibiting optical ($\langle \cos^2 \vartheta_{\mathrm{TDM}} \rangle$) and/or electrical ($\langle \cos \vartheta_{\mathrm{PDM}} \rangle$) anisotropy strongly depends on the conditions used for physical vapor deposition, specifically the ratio between the used substrate temperature $T_S$ and the $T_g$ of the deposited material. Our GSP data of BCPO grown at varied $T_S$ nicely confirm this model (Fig. 5), as the GSP slope vanishes above a critical value of $T_S/T_g \approx 0.9$.

However, for understanding the largely different degrees of alignment $\Lambda$ of the four phosphine oxides grown as films at room temperature (see Table 1), their $T_g$'s cannot be used as a predictor. This rather requires to consider the specific driving forces and the differences in molecular structures. We may speculate that the observed saturation (and even decrease) of the SOP at the lowest substrate temperature might also be related to the surface equilibration model, since molecular motion at the film surface will become insufficient at too low temperature.

# Conclusion

We investigated the influence of molecular design and film-growth conditions on spontaneous orientation polarization in four different phosphine oxides. All of them show SOP, while being optically almost isotropic. Due to the high molecular flexibility different stable conformers exist with diverse easy axes in their electronic polarizability. Consequently, even a high degree of alignment of one molecular axis (here the PO group) can result in vanishing birefringence because of averaging over all different molecular configurations.

Regarding their GSP slopes as well as their degree of PDM alignment, BCPO and POPy2 clearly stand out. Both exhibit just one single PO group resulting in conformers with similar PDM magnitude and direction. Moreover, with their small PDM and just one distinct interaction center positioned at the oxygen atom of the PO group, exceptionally high degrees of SOP were achieved. By contrast, molecules, like DPEPO and PO9, with two PO groups have many conformers with large PDM on average and two competing interaction centers, which results only in a much smaller GSP slope and lower degree of PDM alignment.

By varying the film-growth conditions we are able to tune the degree of PDM alignment over a broad range. Heating the substrate causes the molecular packing to become more random, resulting in vanishing SOP close to the $T_g$ of the material. On the other hand, the GSP slope can be doubled and reaches a record value of almost $300\,\mathrm{mV\,nm^{-1}}$ for BCPO films grown at low temperature. Additionally, by dilution in a non-polar host in combination with substrate cooling an almost complete alignment of the molecules can be achieved, with more than $84\,\%$ of all PDMs in BCPO standing upright on the sample surface on average.

# Methods

## Sample preparation and Kelvin probe (KP) measurements

Films were grown by physical vapor deposition either on silicon substrates (with a natural oxide layer) or on boron crown glass (BK7) substrates in a high vacuum chamber at a base pressure below $1 \times 10^{-6}$ mbar. For the investigation of the different phosphine oxides the deposition rate was kept constant at $1 \, \text{Å} \, \text{s}^{-1}$. The further investigations of BCPO were performed under the specific conditions given in the respective results section.

Kelvin probe measurements were performed in-situ in an ante-chamber attached to the deposition chamber without breaking the vacuum. In the whole process of deposition and measuring, the sample was kept at its respective substrate temperature $T_S$, while light irradiation into the chamber was minimized. A calibrated Kelvin probe system KP6500 from McAllister Technical Services was used to measure the contact potential difference between the vibrating metal tip and the sample. The time between deposition and measurement was kept similar for each deposited layer of a specific sample.

## Impedance spectroscopy (IS)

For impedance spectroscopy, a Solartron Impedance Analyzer model SI 1260 combined with a Dielectric Interface model SI 1296 was used. A constant DC bias superimposed with a small AC voltage (50 mV rms) has been applied to the samples, which consisted of a heterolayer structure of a non-polar hole transport material and the polar SOP material as electron transport layer. The DC bias was swept from reverse to forward while measuring the capacitance of the device. To determine the surface charge density, the transition voltage $V_{tr}$ and the built-in voltage $V_{bi}$ were estimated as the first and second increase in capacitance from the measurements. The polarization charge $\sigma$ is obtained from the capacitance of the polar electron transport layer, $C_{ETL}$, in the range between $V_{tr}$ and $V_{bi}$. For further details of the method we refer to the SI, Section S2, and previous work by our group.[2,60]

## Variable angle spectroscopic ellipsometry (VASE)

Wavelength dependent optical constants and film-thickness were obtained by variable-angle spectroscopic ellipsometry at different angles from $40°$ to $70°$ in the spectral range of 280 nm to 850 nm. The measurements were carried out on a Sentech SE850 instrument and analyzed by the software SpectraRay3, also supplied by Sentech. Further details are given in the SI, Section S3.

## Angular dependent photoluminescence (ADPL)

Angular dependent photoluminescence was used to determine the orientation of the molecular transition dipole moments. A commercial instrument (Phelos by Fluxim AG) was used to collect the emission spectra at different angles. For excitation a 275 nm LED was used and the emitted light was measured under s- and p-polarization relative to the detection plane. The gained raw PL spectra were analyzed by our in-house software. Further details are given in the SI, Section S6.

## DFT calculations

Molecular properties like the permanent dipole moment, polarizability tensor and the transition dipole moment were obtained from density functional theory calculations. We used the commercial Schrödinger Materials Science Suite.[61] First, we searched for possible conformations of each molecule with a Macro Model,[62] which is based on a torsional sampling method. Afterwards the conformations were optimized by Jaguar[63] to gain the energetically optimized structures in the electronic ground state. This was done by performing density functional theory calculations with the basis set 6-31G** and the functional B3LYP. From these optimized structures the electrical dipole moment and the polarizability eigenvectors were calculated. For DPEPO and PO9 additional molecular dynamics simulations were performed to gain a more accurate PDM on average by integrating over the distribution of conformers of a simple box of molecules. Additionally, by further performing time-dependent DFT with a Tamm-Dancoff approximation on the optimized structures, the transition dipole moment of BCPO was obtained.

**Acknowledgement** This work was funded by Deutsche Forschungsgemeinschaft (DFG) under grants BR1728/20-3 (project no. 341263954) and BR1728/22-1 (project no. 432420985).

# Supporting Information Available

- Overview of GSP data

- Impedance spectroscopy of different POs

- Ellipsometry of different POs

- Kelvin probe of BCPO

- Ellipsometry of BCPO

- Angular dependent photoluminescence of BCPO

- DFT simulations of BCPO and POPy2

# Supporting Information

# —

# Controlling Spontaneous Orientation Polarization in Organic Semiconductors - The Case of Phosphine Oxides

Albin Cakaj, Markus Schmid, Alexander Hofmann, Wolfgang Brütting[*]

Institute of Physics, University of Augsburg, 86135 Augsburg, Germany

September 6, 2023

## S1  Overview of GSP data

The maximum polarization charge density of an SOP material is given by the PDM magnitude $p$ of the involved dipolar molecules and their number density $n$ in the solid film:

$$\sigma^{\mathrm{max}} = p\,n = p\,\varrho\,\frac{N_A}{M_W}\,. \tag{1}$$

The latter equality follows, if $n$ is expressed by the mass density $\varrho$, the molecular weight $M_W$ and Avogadro's number $N_A$.

Taking typical values of density and molecular weight, one can estimate the maximum possible polarization value per dipole moment:

$$\frac{\sigma^{\mathrm{max}}}{p} = \varrho\,\frac{N_A}{M_W} \tag{2}$$

For Alq$_3$ we have: $p = 4.55\,\mathrm{D}$, $\varrho = 1.3\,\mathrm{g/cm^3}$ and $M_W = 459\,\mathrm{g/mol}$ so that $\frac{\sigma^{\mathrm{max}}}{p} = 5.6\,\mathrm{mC\,m^{-2}\,D^{-1}}$; OXD-7 gives $\frac{\sigma^{\mathrm{max}}}{p} = 4.5\,\mathrm{mC\,m^{-2}\,D^{-1}}$.

Thus, as a typical value of many organic semiconductors we can use:

$$\frac{\sigma^{\mathrm{max}}}{p} \approx 5\,\mathrm{mC\,m^{-2}\,D^{-1}}\,. \tag{3}$$

---

[*]E-mail: Wolfgang.Bruetting@physik.uni-augsburg.de

Accordingly, the maximum possible GSP slope $m_{\text{GSP}}$ per dipole moment can be predicted as well:

$$\frac{m_{\text{GSP}}^{\text{max}}}{p} = \frac{\sigma^{\text{max}}}{p} \cdot \frac{1}{\varepsilon_0 \, \varepsilon_r} \tag{4}$$

With an $\varepsilon_r = 3$ this yields a typical value of

$$\frac{m_{\text{GSP}}^{\text{max}}}{p} \approx 190 \, \text{mV} \, \text{nm}^{-1} \, \text{D}^{-1} \, . \tag{5}$$

The actual GSP slopes, however, are typically much smaller as shown in the overview graph of Fig. 2 in the manuscript.

Therefore one can define the degree of alignment:

$$\Lambda = \frac{\sigma^{\text{meas}}}{\sigma^{\text{max}}} \quad \text{or} \quad \Lambda = \frac{m_{\text{GSP}}^{\text{meas}}}{m_{\text{GSP}}^{\text{max}}} \tag{6}$$

Note that these two values need not be identical, because IS and KP can yield slightly different results. The shaded area in Fig. 2 of the manuscript indicates the range of GSP slopes to be expected, if typical values for the degree of alignment $\Lambda$ between 4% ($\text{Alq}_3$) and 12% (OXD-7) are used.

Table 1 shows measured data for the materials investigated by our group in recent years, which are included in Fig. 2 of the manuscript. The data either come from impedance spectroscopy (IS) or Kelvin probe (KP).

Table 2 gives further details of the material properties and the calculated maximum polarization charge and GSP slope that were used to obtain the degree of alignment given in Table 1.

| Material | $p$ (D) | $\sigma$ (mC/m²) | $m_{\mathrm{GSP}}$ (mV/nm) | $\Lambda$ (%) | Ref. |
|---|---|---|---|---|---|
| CBP | 0.0 | — | −1.95 | — | [1] |
| NPB | 0.6 | — | 1.12 | 1.3 | [1] |
| Alq$_3$ | 4.55 | 1.1 - 1.7 | — | 4.2 - 6.5 | [2, 3, 4] |
|  |  |  | 33 - 48 | 3.6 - 5.2 | [5, 1] |
| OXD-7 | 5.4 | 2.6 - 3.5 | — | 10.4 - 14.1 | [6, 1] |
|  |  | — | 78 - 114 | 8.0 - 11.7 | [1, 7] |
| TPBi | 6.3 | — | 69 | 7.4 | [7] |
| Ir(ppy)$_3$ | 6.1 | 0.0 | — | 0.0 | [8] |
| Ir(ppy)$_2$(acac) | 2.0 | 3.3 | — | 32.7 | [8] |
|  |  | — | 55 | 14.4 | [1] |
| mer-Ir(dbfmi)$_3$ | 5.4 | — | −42 - −55 | 6.8 - 8.9 | [9] |
| mCBP-CN | 3.7 | 1.7 | — | 9.9 | [10] |
| DPEPO | 7.1 | 1.6 - 1.7 | — | 5.4 - 5.8 | [9, 10] |
| PO9 | 7.8 | 1.2 | — | 4.2 | [10] |
| BCPO | 3.5 | 4.3 | — | 31.4 | [10] |

Table 1: SOP of materials investigated by our group and their respective polarization charge density (measured by IS) or GSP slope (measured by KP), which are shown in Fig. 2 of the manuscript. The degree of alignment was calculated according to the respective maximum values of $\sigma$ and $m_{\mathrm{GSP}}$ given in Table 2.

| Material | $p$ (D) | $\rho$ (g/cm³) | $M$ (g/mol) | $\varepsilon_r$ | $m_{\mathrm{GSP}}^{\mathrm{max}}$ (mV/nm) | $\sigma^{\mathrm{max}}$ (mC/m²) |
|---|---|---|---|---|---|---|
| CBP | 0.0 | 1.2 | 485 | 2.7 | 0 | 0 |
| NPB | 0.6 | 1.2 | 589 | 3.3 | 83.5 | 2.44 |
| Alq$_3$ | 4.55 | 1.3 | 459 | 3.2 | 920 | 26.1 |
| OXD-7 | 5.4 | 1.1 | 479 | 2.9 | 971 | 24.9 |
| TPBi | 6.3 | 1.3 | 655 | 3 * | 932 | 24.8 |
| Ir(ppy)$_3$ | 6.1 | 1.5 | 655 | 3 * | 1042 | 27.7 |
| Ir(ppy)$_2$(acac) | 2.0 | 1.5 | 600 | 3 * | 380 | 10.1 |
| mer-Ir(dbfmi) | 5.4 | 1.3 | 934 | 3 * | −579 | −15.4 |
| mCBP-CN | 3.7 | 1.2 | 510 | 3 * | 649 | 17.2 |
| DPEPO | 7.1 | 1.2 | 571 | 3 * | 1113 | 29.5 |
| PO9 | 7.8 | 1.2 | 643 | 3 * | 1086 | 28.8 |
| BCPO | 3.5 | 1.2 | 609 | 3 * | 514 | 13.7 |

Table 2: Properties used for the calculation of the degree of alignment in Table 1. (* For simplicity $\varepsilon_r = 3$ was used in these cases.)

## S2 Impedance spectroscopy of different phosphine oxides

IS was performed on simple heterolayer devices. They consisted of PEDOT:PSS (Al4083) as hole injection layer (HIL), 70 nm NPB as non-polar hole transporting layer (HTL), 70 nm of the phosphine oxide as electron transporting layer (ETL) and a cathode with 15 nm calcium and 100 nm aluminum.

In Figure 1a exemplary impedance spectroscopy measurements can be seen. The surface charge density $\sigma$ was calculated from the following equation:

$$\sigma = \frac{C_{ETL}}{A}(V_t - V_{bi}),\tag{7}$$

where $C_{ETL}$ is the plateau value of the capacitance after hole accumulation, which starts at a threshold voltage $V_t$. For the built-in voltage $V_{bi}$ a value of 2.4 V, corresponding to the work function difference of the used contacts, was used for the calculation of $\sigma$. Note that the obtained values are negative, in this case, because IS probes the polarization charge at the bottom interface of the SOP layer. The pixel area $A$ was measured with a microscope.

Furthermore, for DPEPO, PO9 and BCPO, we also performed a layer thickness variation so that the polarization charge is obtained from the slope of a linear dependence (Fig. 1b):

$$V_t = \sigma \frac{A}{C_{ETL}} + V_{bi} = \sigma \frac{d_{ETL}}{\varepsilon_0 \varepsilon_r} + V_{bi}\tag{8}$$

and the built-in voltage from the intercept with the vertical axis. The latter equality can be used, to obtain the relative dielectric constant $\varepsilon_r$ as well. Both methods yield consisted values of $\sigma$, which are plotted in Fig. 4 of the manuscript (as absolute values).

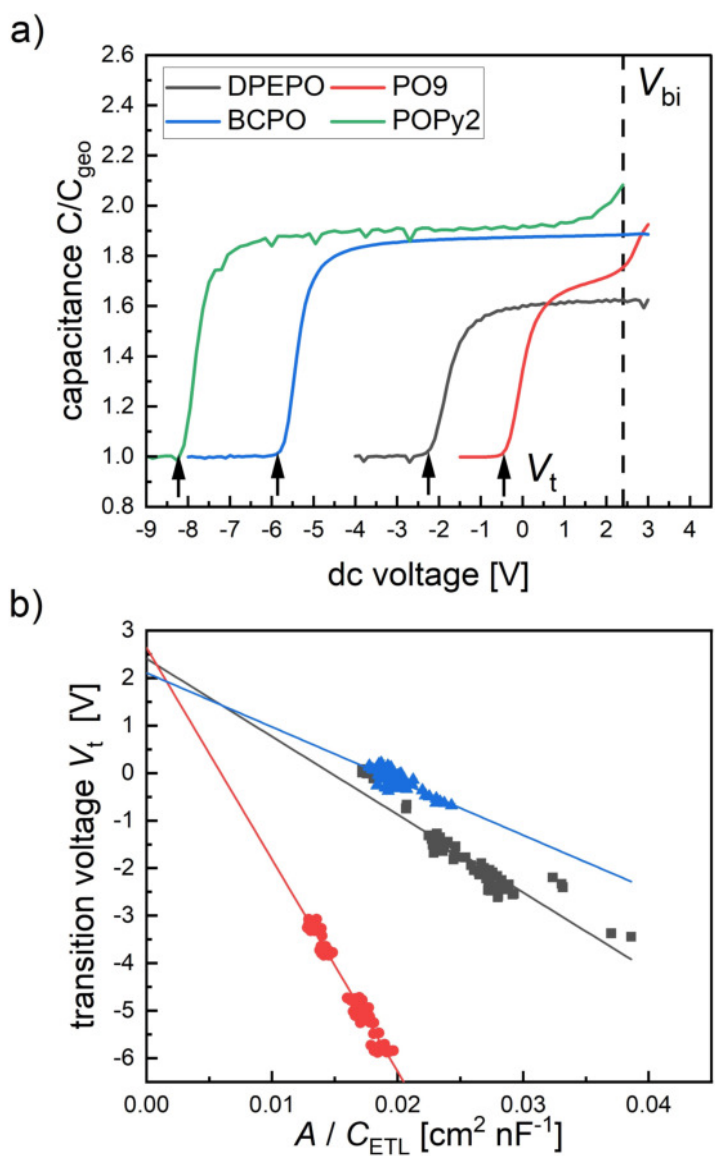

Figure 1: a) Exemplary impedance measurements of the four investigated phosphine oxides. b) Layer thickness variation for three of the investigated phosphine oxides namely DPEPO, PO9 and BCPO.

## S3 Ellipsometry of different phosphine oxides

VASE is based on a change of the polarization state of light from linear to elliptic upon reflection from the sample. This is expressed by the two ellipsometric angles $\Delta$ and $\Psi$, as shown in Fig. 2. Fitting these curves requires the use of appropriate optical models, which in turn allows to obtain the optical constants ($n \& k$) as well as the film thickness. Additionally by performing the measurement under different angles we can probe, whether a film is birefringent or optically isotropic.

As a first step, a simple Cauchy model describing transparent films was used in the region away from the respective absorption in order to extract the film thickness. This was done for all samples measured by Kelvin probe to verify the total thickness after measurement. Samples exhibiting noticeable deviations to their nominal thickness were corrected with respect to the CPD slopes from KP measurements. This was specifically the case for the films prepared at higher evaporation rates.

For the extraction of both thickness and optical constants, either a Cauchy or an oscillator model (Tauc-Lorentz or Drude-Lorentz) was used. Additionally to check if a film is birefringent, an additional anisotropic layer was introduced in the model. It contains two separate layers describing each the optical properties in-plane and out-of-plane with respect to the surface normal. Only if the fit with the additional complexity of such a model is significantly better, can the film be considered as truly anisotropic.

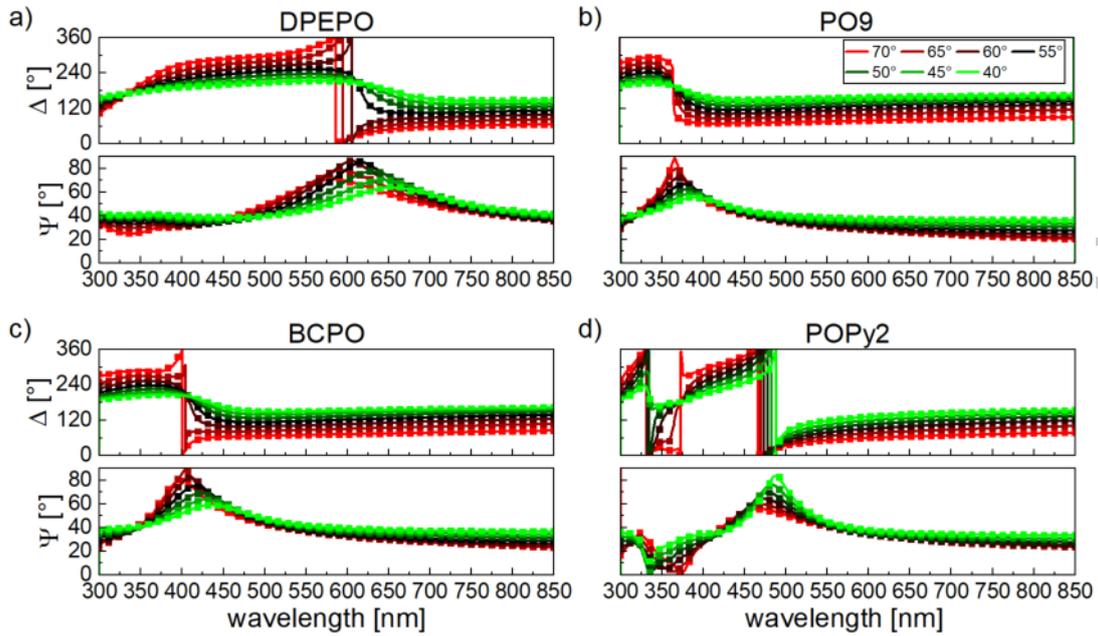

Figure 2: Raw data and fits by variable angle spectroscopic ellipsometry of all investigated phosphine oxides. In the fitting procedure different layer models were used. a) DPEPO: Cauchy model, b) PO9: Drude-Lorentz oscillator model. For c) BCPO and d) POPy2 a Tauc-Lorentz oscillator model was used with a sum of oscillators.

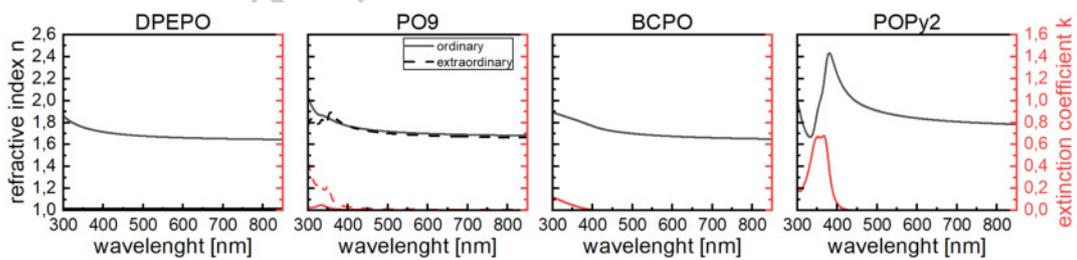

Figure 3: Resulting optical constants of all four investigated phosphine oxides. Only PO9 showed a noticeable anisotropy in the optical constants.

## S4 Kelvin probe measurements of BCPO

### S4.1 Variation of substrate temperature

Contact potential difference (CPD) between the vibrating KP head and the sample was measured on films with increasing thickness. The step size for film growth was between 5 and 10 nm, depending on the magnitude of the CPD. The KP instrument was limited to a maximum of 10 V in CPD detection.

GSP slopes were obtained from linear fits of the CPD with increasing film thickness, whereby the value of the bare substrate was excluded from the fit.

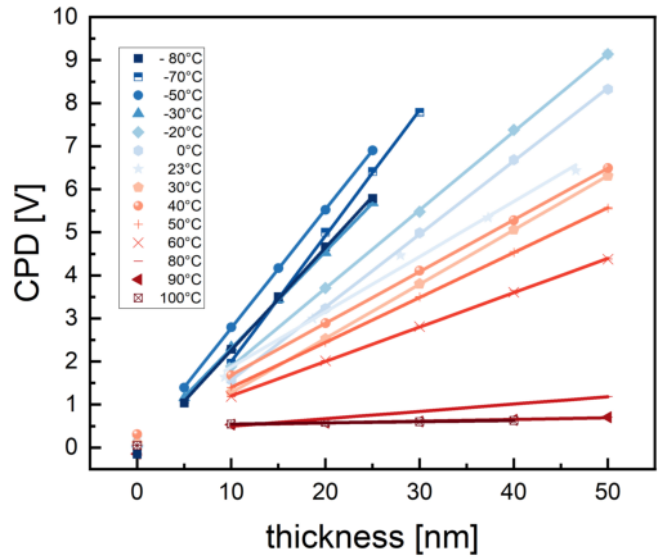

Figure 4: Raw KP data of BCPO films grown at different substrate temperatures.

**S4.2  Variation of evaporation rate**

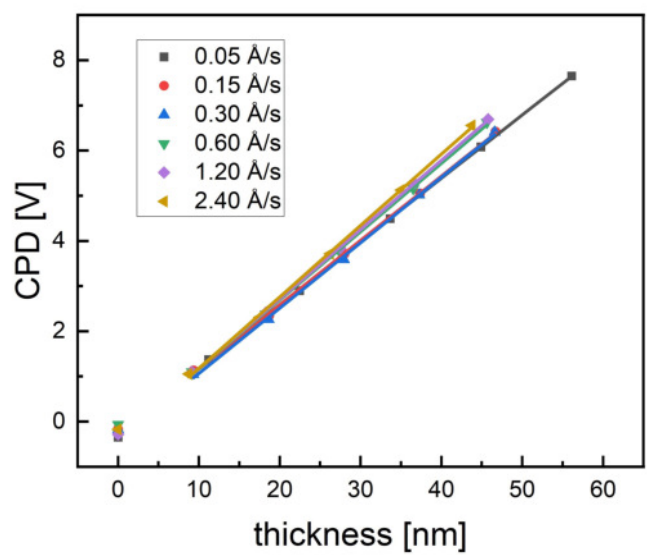

Figure 5: Raw KP data of BCPO films grown at different evaporation rates. The substrate temperature was kept at room temperature.

## S4.3  BCPO doped in CBP as guest-host system

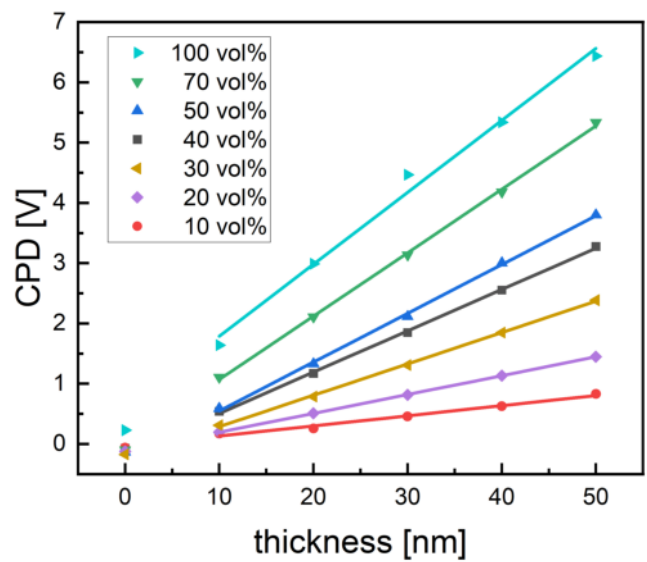

Figure 6: Raw KP data of BCPO doped in CBP at different volume concentrations. The temperature was kept at room temperature for all samples.

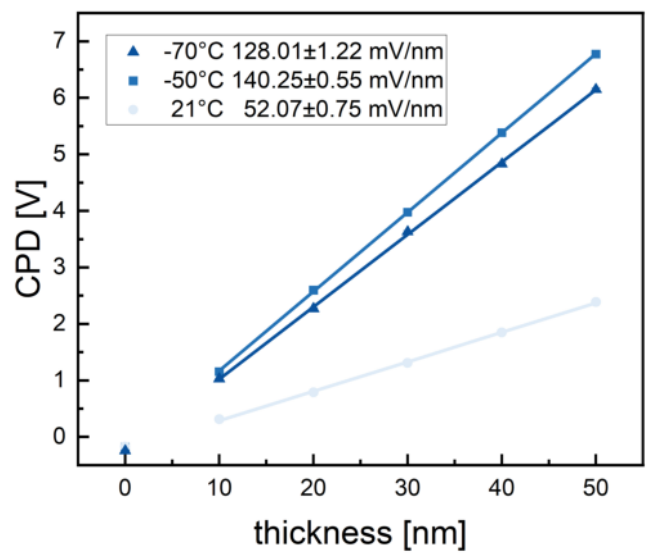

Figure 7: Raw KP data of films with BCPO doped in CBP at 30% volume concentrations at different substrate temperatures. The evaporation rate was kept constant at 0.3 Å/s for all samples.

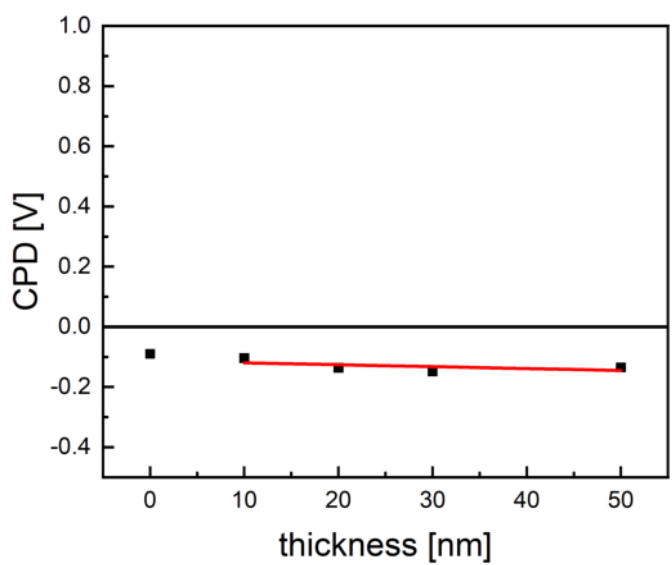

Figure 8: Raw KP data of neat CBP films grown at room temperature. CBP has an almost negligible GSP slope of $-0.64\,\mathrm{mV/nm}$.

# S5 Ellipsometry of BCPO

## S5.1 Variation of substrate temperature

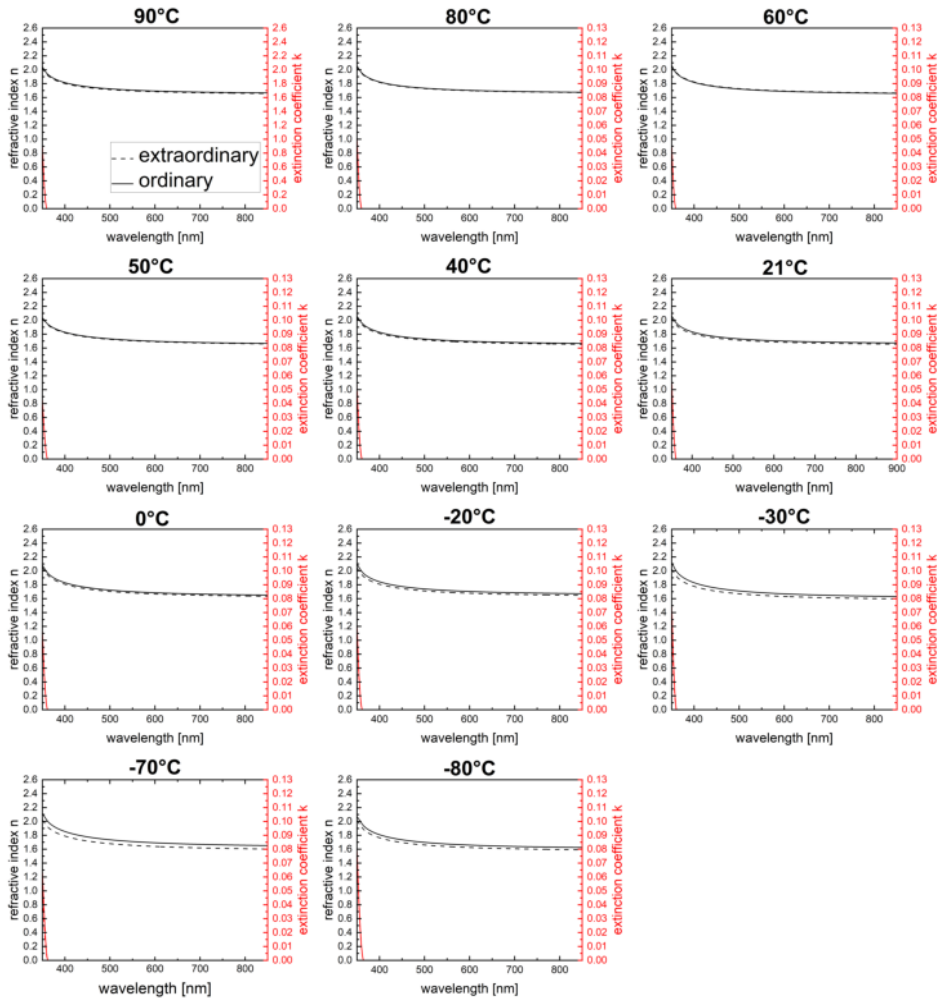

Figure 9: Optical constants of BCPO films prepared at different substrate temperatures. For fitting a Tauc-Lorentz-oscillator model with only one oscillator was used in a uniaxial model. The $\Delta n$ values shown in Fig. 5b of the main text were taken at a wavelength of 632.8 nm.

## S5.2 Variation of evaporation rate

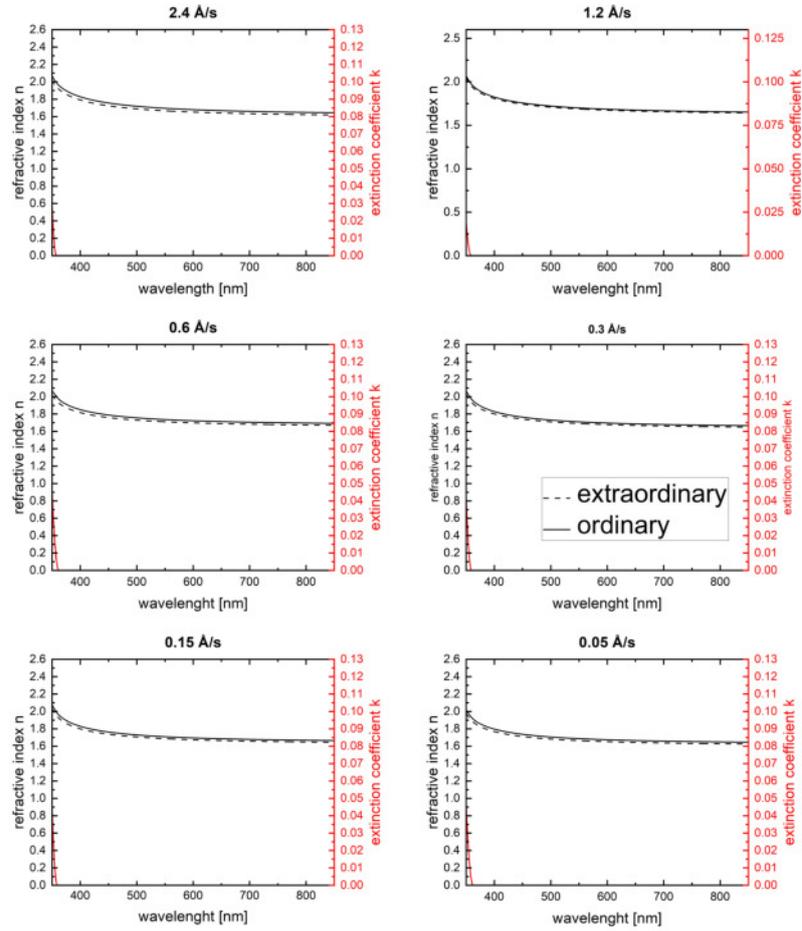

Figure 10: Optical constants of BCPO films prepared at different evaporation rates. For fitting a Tauc-Lorentz oscillator model with only one oscillator was used in a uniaxial model.

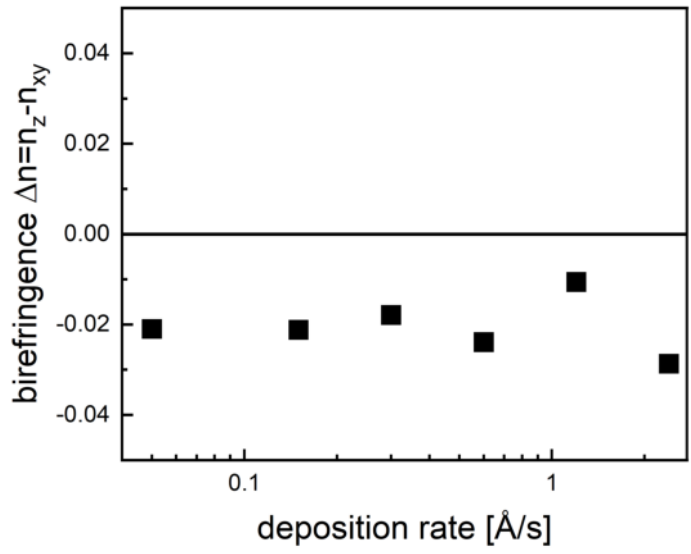

Figure 11: Resulting birefringence of BCPO films prepared at different evaporation rates. The values were determined at a wavelength of 632.8 nm.

# S6 Angular dependent photoluminescence measurements of BCPO

The PL of BCPO films has an emission peak at about 380 nm. However, due to the need for a polarizer for measuring angular dependent PL, which cuts off below 400 nm (see Figure 12), we have performed the optical orientation analysis right at this wavelength. The resulting angular dependent PL curves and the corresponding fits with the obtained optical orientation parameter $\Theta$ for films grown at different substrate temperature are shown in Fig. 13.

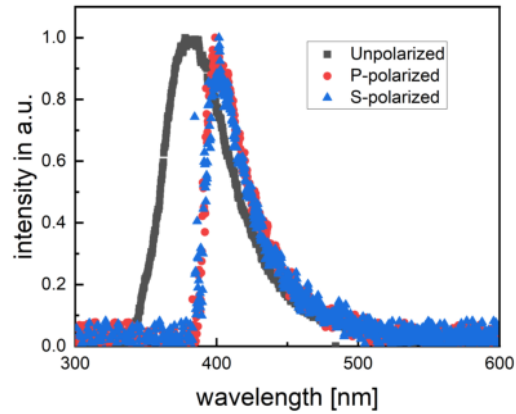

Figure 12: Photoluminescence spectra of BCPO without (black) and with (red/blue) polarizer.

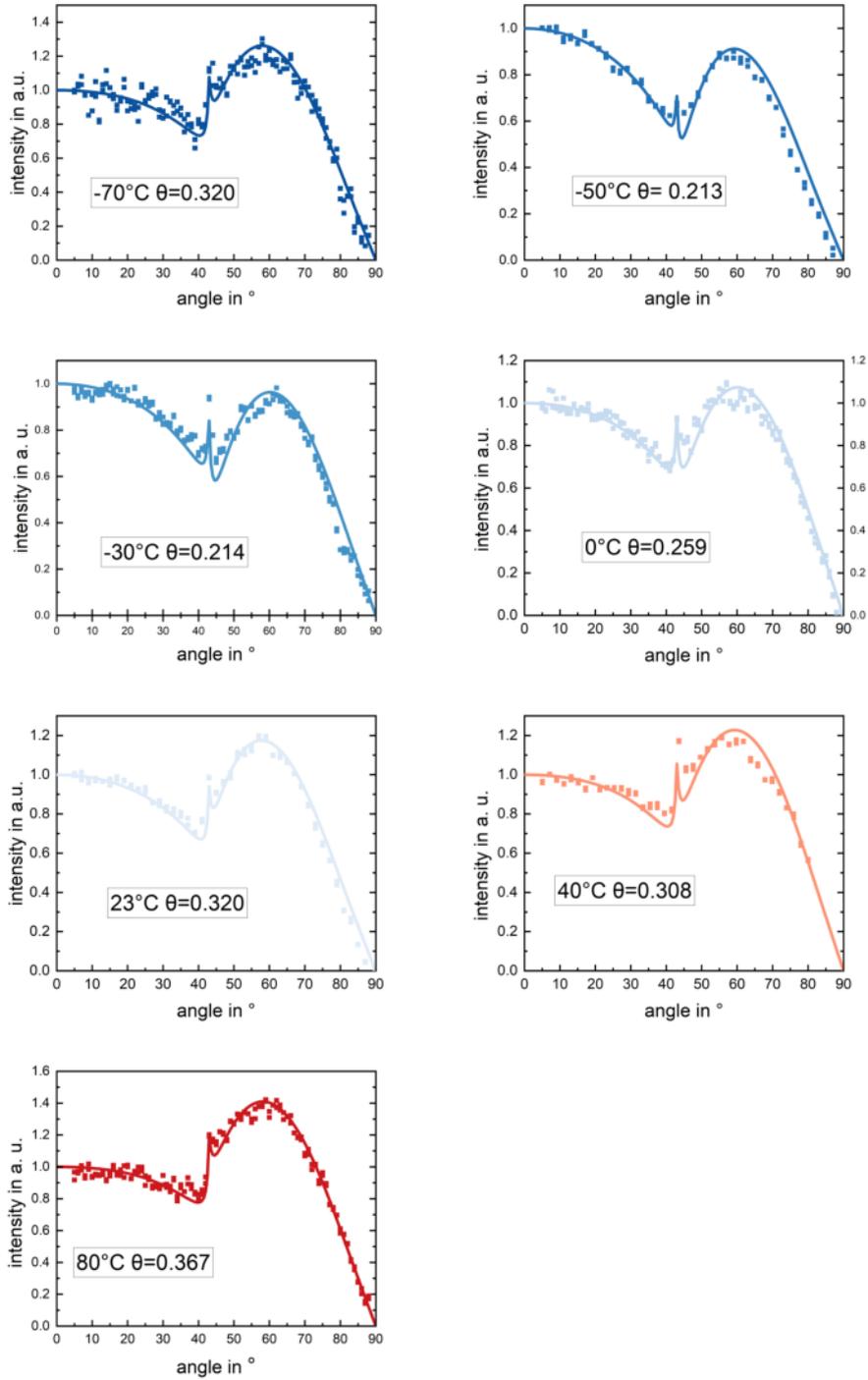

Figure 13: ADPL measurements and optical orientation fits of BCPO films grown at different substrate temperatures (taken at a wavelength of 400 nm).

## S7 DFT simulations of BCPO and POPy2

DFT simulations were used to perform geometry optimization of the molecules BCPO and POPy2 and to obtain the directions of their respective PDMs for some exemplary structures with high Boltzmann weight.

In case of BCPO, we also calculated the TDM from TD-DFT, which is used to discuss the relation between SOP and optical orientation from VASE and ADPL. This was not necessary for POPy2, because no optical measurements were performed for this material.

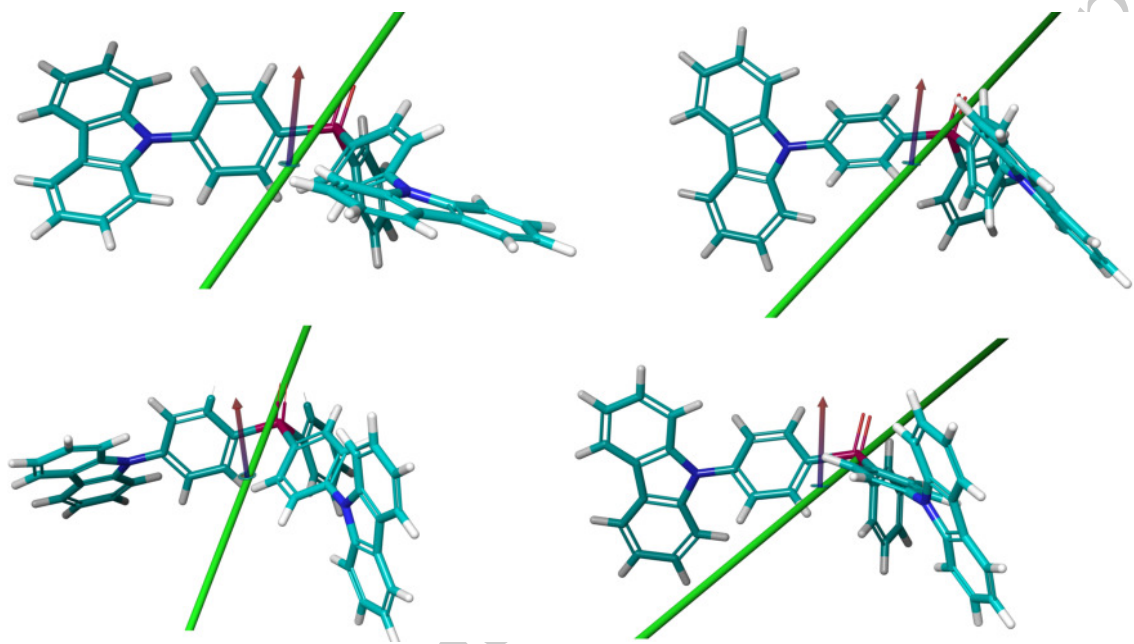

Figure 14: Exemplary structures of BCPO obtained from (TD)-DFT. PDM is shown in red (with values of 3.3 - 3.5 D), TDM in green (arbitrary units).

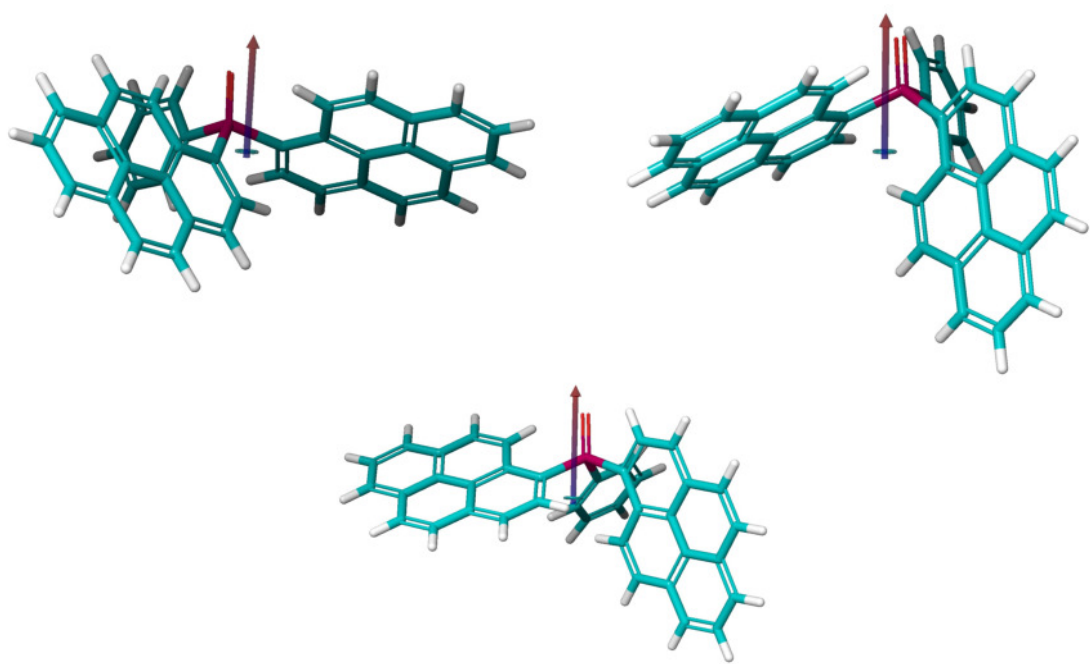

Figure 15: Exemplary structures of POPy2 obtained from DFT. PDM is shown in red (3.6 - 4.2 D).